\documentclass[12pt]{lcs}
\usepackage{psfig}
\newcommand{\double}{\baselineskip 2 \baselineskip}

\begin{document}

\double

\title{A topology visualisation tool for large-scale
communications networks}

\author{Yuchun Guo, Changjia Chen and Shi Zhou}

\date{}

\maketitle

A visualisation tool is presented to facilitate the study on
large-scale communications networks. This tool provides a simple and
effective way to summarise the topology of a complex network at a
coarse level.

\paragraph{Introduction} It is relevant to study the  topology of
information and communications networks, such as the Internet, the
World Wide Web and peer-to-peer (P2P) overlay networks, because
structure fundamentally affects functions. These networks contain
thousands or even millions of connections and their topologies are
usually characterised by statistics~\cite{pastor04, newman06}. Here
we introduce a simple tool to visualise a network's connectivity.

\paragraph{Methodology} In graph theory degree is defined as the number
of links a node has. We sort network nodes in a list of decreasing
degrees, i.e.~at the top position is the best-connected node and next
is the second best-connected node. For each group of nodes that have
the same degree, we rearrange the nodes in the decreasing order of
their neighbours largest degree; for those having the same neighbours
largest degree as well, we sort their neighbours second-largest
degree; and this process continues until all neighbours degrees have
been considered. We then assign each node an index according to its
position in the sorted list.

If a network contains $N$ nodes, the network's connectivity
information can be represented as an $N \times N$ adjacency matrix,
in which entry $a_{ij}$ is the number of links connecting between
nodes with the sorted indexes of $i$ and $j$, where $i, j=1,2,3...N$.
For a simple undirected graph (i.e.~no self-loop and no multiple
links between a pair of nodes), the adjacency matrix is a symmetric
$(0,1)$-matrix with zeros on its diagonal. Then the network's
topology information can be visualised  as a bitmap of sorted
adjacency matrix (BOSAM). One may consider the neighbours largest
degree as a measure of the `cohesion' force a node receives from the
network core of high-degree nodes. Then an alternative BOSAM can be
obtained by sorting the neighbours smallest degree, i.e.~a measure of
the `radiation' force the node receives from the peripheral
low-degree nodes.

\paragraph{Results} Figure~1(a) shows BOSAM of the Internet topology at
the autonomous systems (AS) level collected by
CAIDA~\cite{mahadevan05b}, on which black pixels are densely
concentrated along the top and the left borders where $i<0.5k$ or
$j<0.5k$. This indicates that the Internet AS graph exhibits the
following properties: (1) power-law degree
distribution~\cite{faloutsos99}, which means a small number of nodes
have very large numbers of links whereas the majority of nodes have
only a few links; (2) negative degree-degree
correlation~\cite{vazquez03}, which means low-degree nodes tend to
connect to high-degree nodes and vice versa; and (3) rich-club
phenomenon~\cite{zhou04a}, which describes the fact that high-degree
nodes, or rich nodes, are tightly interconnected with themselves.
These properties explain why the Internet is so `small' that the
average shortest path between a pair of nodes is only 3.12 hops:
while the rich nodes know each other very well and collectively
function as a super traffic hub for the network, the majority of the
nodes, peripheral low-degree nodes, are always near the rich-club
core. Figure~1(b), (c) and (d) show that by comparison, the
positive-feedback preference (PFP) Internet model~\cite{zhou04d}
better resembles the Internet AS graph than the Erd\"os-R\'enyi (ER)
random model~\cite{erdos59}  and the Barab\'asi-Albert (BA)
scale-free model~\cite{Barabasi99a}.

The Gnutella peer-to-peer (P2P) file-sharing
network~\cite{stutzbach05} contains 317,592 nodes and 7,396,948
links. Figure~2 shows Gnutella is profoundly different from the
Internet AS graph. Instead it shows similarity to the ER random
model. Figure~2 also reveals that Gnutella exhibits a fractal-like
structure which has not been captured by the usual statistic studies.

\paragraph{Conclusion} Although BOSAM is a coarse representation of a
network's structural properties, we show that this simple tool is
effective in distinguishing network topologies.

\paragraph{Acknowledgment} YG and CC are supported by Chinese NSFC grant
no.~60672069 and SRFDP grant no.~20050004033. SZ is supported by the
UK Nuffield Foundation grant no.~NAL/01125/G.

\vspace{10mm}

Y.~Guo, C.~Chen (\emph{Beijing JiaoTong University, China}) and
S.~Zhou (\emph{University College London, UK})

E-mail: ychguo@bjtu.edu.cn

\newpage

\paragraph{Figure 1} BOSAM of (a) the Internet AS graph collected by CAIDA; (b) the PFP
Internet model; (c) the ER random model; and (d) the BA scale-free
model. All networks have 9,204 nodes and 27,612 links.

\paragraph{Figure 2} BOSAM of (a)~Gnutella P2P network and enlargements
of the top-left corner of (a) at scales of (b) 1:4, (c) 1:16, and (d)
1:64.

\newpage

\begin{figure}
\centerline{\psfig{figure=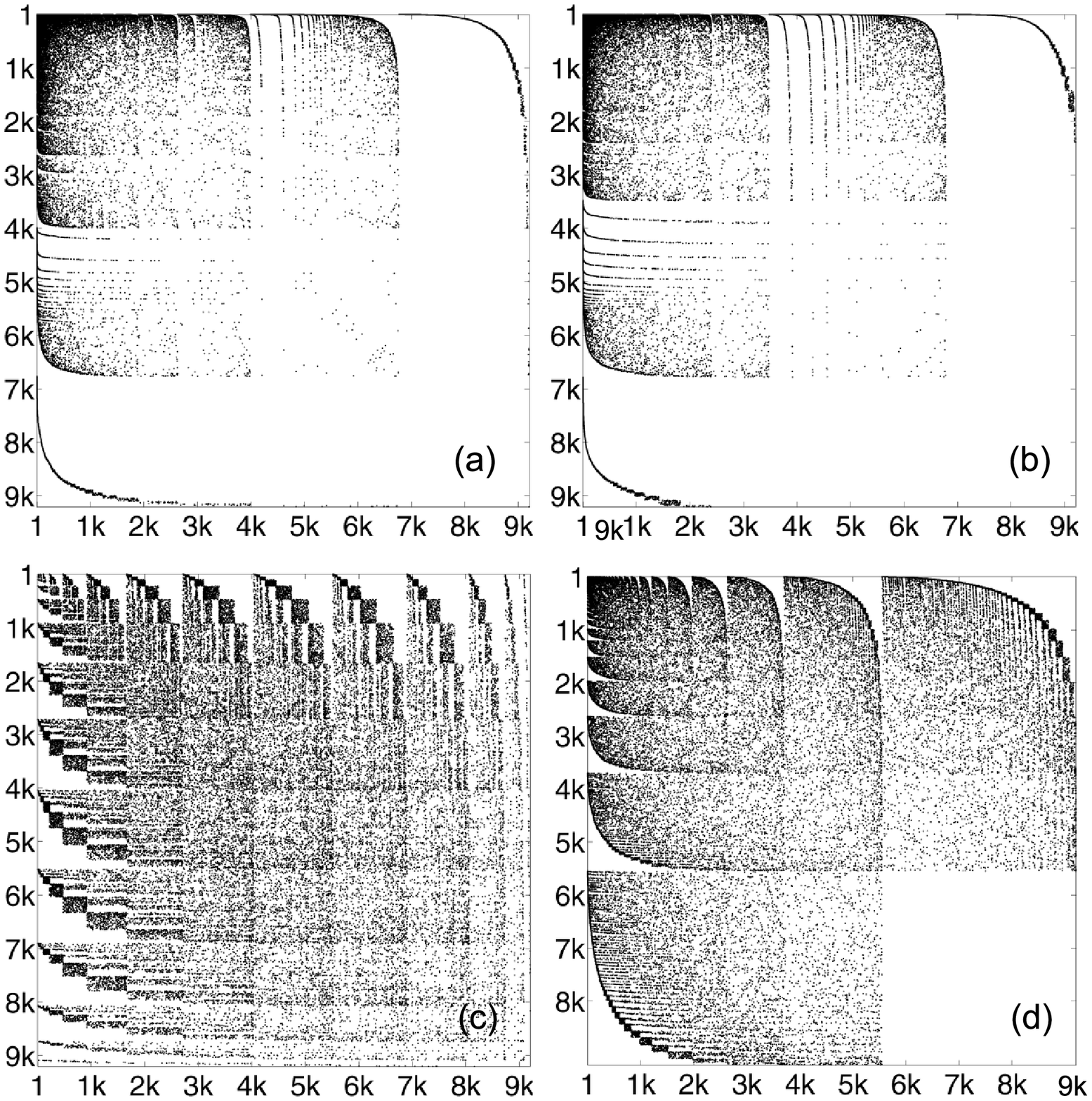,width=12cm}}\caption{BOSAM of
(a) the Internet AS graph collected by CAIDA; (b) the PFP Internet
model; (c) the ER random model; and (d) the BA scale-free model. All
networks have 9,204 nodes and 27,612 links.}
\end{figure}

\newpage

\begin{figure}
\centerline{\psfig{figure=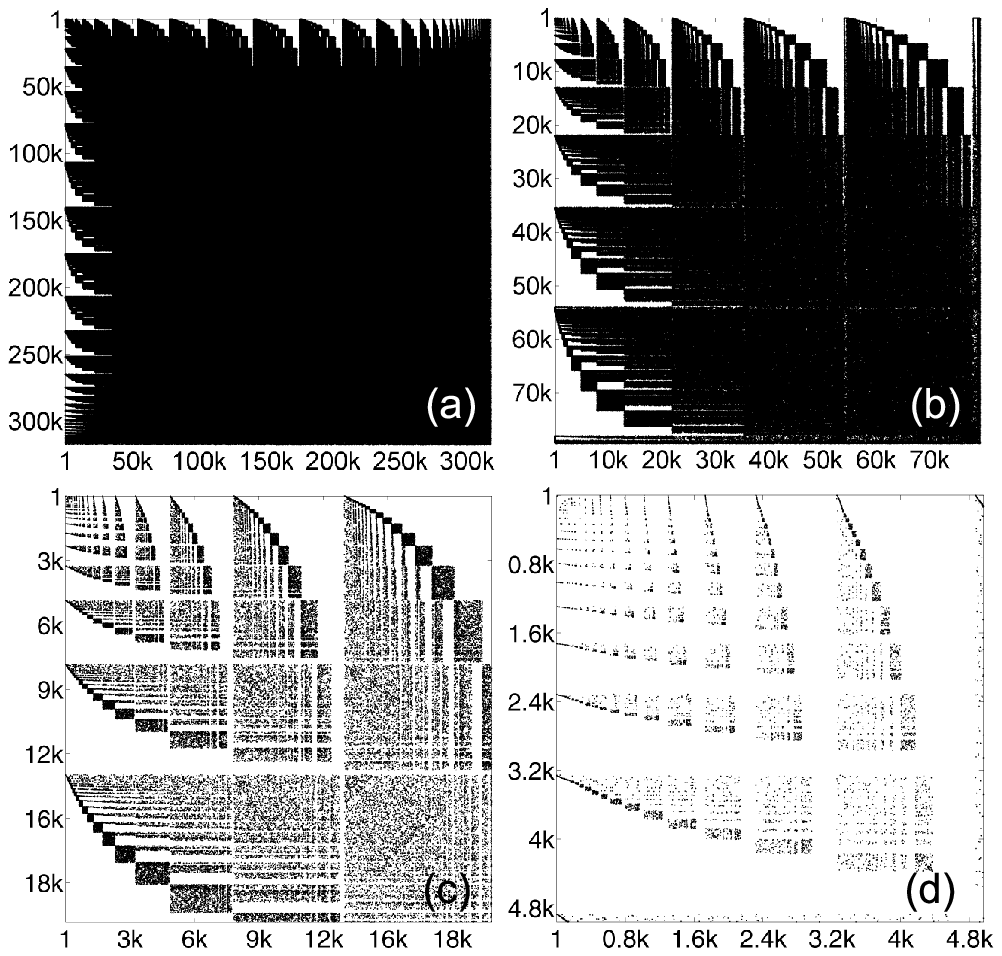,width=12cm}}
\caption{BOSAM of (a)~Gnutella P2P network and enlargements of the
top-left corner of (a) at scales of (b) 1:4, (c) 1:16, and (d) 1:64.}
\end{figure}


\begin{thebibliography}{10}
\providecommand{\url}[1]{#1} \csname url@rmstyle\endcsname
\providecommand{\newblock}{\relax}
\providecommand{\bibinfo}[2]{#2}
\providecommand\BIBentrySTDinterwordspacing{\spaceskip=0pt\relax}
\providecommand\BIBentryALTinterwordstretchfactor{4}
\providecommand\BIBentryALTinterwordspacing{\spaceskip=\fontdimen2\font
plus \BIBentryALTinterwordstretchfactor\fontdimen3\font minus
  \fontdimen4\font\relax}
\providecommand\BIBforeignlanguage[2]{{%
\expandafter\ifx\csname l@#1\endcsname\relax
\typeout{** WARNING: IEEEtran.bst: No hyphenation pattern has been}%
\typeout{** loaded for the language `#1'. Using the pattern for}%
\typeout{** the default language instead.}%
\else \language=\csname l@#1\endcsname \fi #2}}



\bibitem{pastor04}
R.~Pastor-Satorras and A.~Vespignani, \emph{Evolution and Structure
of the
  Internet - A Statistical Physics Approach}.\hskip 1em plus 0.5em minus
  0.4em\relax Cambridge University Press, 2004.


\bibitem{newman06}
M.~Newman, A.-L. Barab\'asi, and D.~Watts, Eds., \emph{The Structure
and
  Dynamics of Networks}.\hskip 1em plus 0.5em minus 0.4em\relax Princeton
  University Press, 2006.

  \bibitem{mahadevan05b}
P.~Mahadevan, D.~Krioukov, M.~Fomenkov, B.~Huffaker,
X.~Dimitropoulos,
  K.~Claffy, and A.~Vahdat, ``The Internet AS-level topology: Three data
  sources and one definitive metric,'' \emph{Comput. Commun. Rev.}, vol.~36,
pp.~17--26, 2006.

    \bibitem{faloutsos99}
M.~Faloutsos, P.~Faloutsos, and C.~Faloutsos, ``On power-law
relationships of
  the Internet topology,'' \emph{Comput. Commun. Rev.}, vol.~29, pp.~251--262,
  1999.

    \bibitem{vazquez03}
A.~Vazquez, R.~P.-S. M.~Boguna, Y.~Moreno, and A.~Vespignani,
``Topology and
  correlations in structured scale-free networks,'' \emph{Phys. Rev. E},
  vol.~67, no.~046111,  2003.

\bibitem{zhou04a}
S.~Zhou and R.~J. Mondrag\'on, ``The rich-club phenomenon in the
Internet
  topology,'' \emph{IEEE Comm. Lett.}, vol.~8, no.~3, pp.~180--182, 2004.

\bibitem{zhou04d}
S.~Zhou and R.~J. Mondrag\'on, ``Accurately modelling the Internet
topology,''
  \emph{Phys. Rev. E}, vol.~70, no.~066108,  2004.

\bibitem{erdos59}
P.~Erd\H{o}s and A.~R\'enyi, ``On random graphs,'' \emph{Publ.
Math.}, vol.~6,  pp.~290--297, 1959.

  \bibitem{Barabasi99a} A.~L. Barab\'asi and
R.~Albert, ``Emergence of scaling in random networks,''
  \emph{Science}, vol.~286, pp.~509--512, 1999.

\bibitem{stutzbach05}
D.~Stutzbach, R.~Rejaie, and S.~Sen, ``Characterizing unstructured
overlay
  topologies in modern p2p file-sharing systems,'' in \emph{Proc.~of the ACM
  Internet Measurement Conference (IMC)}, 2005.


\end{thebibliography}
\end{document}